\documentclass[final,5p,twocolumn,authoryear]{elsarticle}

\usepackage{amssymb}
\usepackage{graphicx,amsmath}
\usepackage{ifthen}
\usepackage{xr}
\usepackage{multirow}
\usepackage{url}
\usepackage{hyperref}
\hypersetup{colorlinks=true,citecolor=blue}
\usepackage{breakurl}

\def\beq{\begin{equation}}
\def\eeq{\end{equation}}
\def\bea{\begin{eqnarray}}
\def\eea{\end{eqnarray}}
\def\nn{\nonumber}

\biboptions{round,comma,sort&compress}

\journal{Energy Policy}

\begin{document}

\begin{frontmatter}


\title{The dynamics of technology diffusion and the impacts of climate policy instruments in the decarbonisation of the global electricity sector}

\author[4cmr]{J.-F. Mercure  \corref{cor1}}
\ead{jm801@cam.ac.uk}

\author[ce]{H. Pollitt}
\author[ce]{U. Chewpreecha}

\author[4cmr]{P. Salas}
\author[4cmr]{A. M. Foley}

\author[ou]{P. B. Holden}
\author[ou]{N. R. Edwards}

\address[4cmr]{Cambridge Centre for Climate Change Mitigation Research (4CMR), Department of Land Economy, University of Cambridge, 19 Silver Street, Cambridge, CB3 1EP, United Kingdom}

\address[ce]{Cambridge Econometrics Ltd, Covent Garden, Cambridge, CB1 2HT, UK}

\address[ou]{Environment, Earth and Ecosystems, Open University, Milton Keynes, UK}

\cortext[cor1]{Corresponding author: Jean-Fran\c{c}ois Mercure}
\fntext[fn1]{Tel: +44 (0) 1223337126, Fax: +44 (0) 1223337130}

\begin{abstract}

This paper presents an analysis of climate policy instruments for the decarbonisation of the global electricity sector in a non-equilibrium economic and technology diffusion perspective. Energy markets are driven by innovation, path-dependent technology choices and diffusion. However, conventional optimisation models lack detail on these aspects and have limited ability to address the effectiveness of policy interventions because they do not represent decision-making. As a result, known effects of technology lock-ins are liable to be underestimated. In contrast, our approach places investor decision-making at the core of the analysis and investigates how it drives the diffusion of low-carbon technology in a highly disaggregated, hybrid, global macroeconometric model, FTT:Power-E3MG. Ten scenarios to 2050 of the electricity sector in 21 regions exploring combinations of electricity policy instruments are analysed, including their climate impacts. We show that in a diffusion and path-dependent perspective, the impact of combinations of policies does not correspond to the sum of impacts of individual instruments: \emph{synergies} exist between policy tools. We argue that the carbon price required to break the current fossil technology lock-in can be much lower when combined with other policies, and that  a 90\% decarbonisation of the electricity sector by 2050 is affordable without early scrapping. 

\end{abstract}

\begin{keyword}

Climate policy \sep Emissions reductions pathways \sep Climate change mitigation \sep  Energy systems modelling

\end{keyword}

\end{frontmatter}



\section{Introduction}

The electricity sector emits 38\% of global energy-related greenhouse gases \citep[GHGs,][]{IEAWEB2012}. Investment planning in the electricity sector is therefore of critical importance to climate-change policy. Electricity production is an energy sector with some of the longest time scales for technological change, requiring particularly careful planning in order to avoid locking in, for many decades, to heavily emitting systems that could commit society to dangerous levels of global warming \citep{IPCCAR5WGIIITS,IPCCAR4WGIIITS}. Meeting emissions targets to prevent warming beyond 2$^{\circ}$C significantly restricts the number of possible pathways of energy sector development \citep{Rogelj2012}. However warming beyond 2$^{\circ}$C is likely to lead to catastrophic consequences for global ecosystems and food chains, with important repercussions for global human welfare \citep{IPCCAR5WGIITS,IPCCAR4WGIITS}. Large reductions in greenhouse gas emissions involve significant amounts of technology substitution, most likely large scale \emph{socio-technical transitions} \citep[as defined by][]{Geels2002}.


Results of techno-economic studies of climate change mitigation depend strongly on assumptions made concerning technology dynamics \citep{Loschel2002}. The majority of studies of energy systems are made using either bottom-up cost-optimisation, or top-down general equilibrium utility optimisation (equilibrium) computational models, or a combination of both. In these models however, dynamics result mostly from the assumptions about optimisation that underpin the modelling approaches. 

In stark contrast with more traditional optimisation-based approaches, this work proposes a new modelling paradigm based exclusively on non-equilibrium dynamics to simulate the impacts of specific policy frameworks, through the economy, onto the environment. We present an analysis of the global electricity sector with high resolution simulations of technology diffusion dynamics, using the `Future Technology Transformations' framework (FTT:Power),\footnote{Future Technology Transformations in the Power sector \citep{Mercure2012}, \burlalt{http://www.4cmr.group.cam.ac.uk/research/esm}{www.4cmr.group.cam.ac.uk/research/esm}.} coupled with non-equilibrium macroeconomics (E3MG),\footnote{The Energy-Economy-Environment Model at the Global level \citep[E3MG/E3ME][]{Barker2010}, \burlalt{http://www.e3mgmodel.com}{www.e3mgmodel.com}, \burlalt{http://www.e3me.com}{www.e3me.com}. E3MG and E3ME are variants of the same model with different regional and sectoral classifications/resolution.} and environmental impacts derived by combining emulators of the climate system (PLASIM-ENTSem) and the carbon cycle  (GENIEem).\footnote{Planet Simulator - Efficient Numerical Terrestrial Scheme - Emulator \citep[PLASIM-ENTSem,][]{Holden2013b}, Grid Enabled Integrated Earth systems model - Emulator \citep[GENIEem,][]{Holden2013}. }

Uncoordinated technology dynamics are modelled at the level of \emph{diverse} profit-seeking investor decisions incentivised by policy under \emph{bounded rationality}, as opposed to system level optimisation. This setup enables us to explore the outcomes of particular energy policy tools for technology diffusion, electricity generation, global emissions, climate change and macroeconomic change. The connection of a diffusion framework to a non-equilibrium model of the global economy opens a very rich world of macroeconomic dynamics and technological change where the impacts of energy policy reveal complex interactions between the energy sector and the economy. 

Ten scenarios of the future global power sector up to 2050 are presented, creating a storyline to provide insight for  the construction of effective comprehensive energy policy portfolios in the context of non-equilibrium dynamics. Going beyond carbon pricing only and considering other policies that could help trigger the diffusion of new technologies, particular combinations are found to feature mutual synergies that provide suitable environments for fast electricity sector decarbonisation: up to 90\% by 2050. Macroeconomic dynamics in these scenarios are summarised. High resolution scenario data are accessible on our website at \burlalt{http://www.4cmr.group.cam.ac.uk/research/FTT/fttviewer}{www.4cmr.group.cam.ac.uk/research/FTT/fttviewer}, where costs, electricity generation and emissions can be explored in 21 world regions and 24 technologies. 

Decarbonisation involves the positive externality associated to the global accumulation of knowledge and experience in scaling up, deploying and using new power technologies. A classic collective action problem emerges: learning cost reductions for new technologies may only become significant and enable cost-effective diffusion when most nations of the World demonstrate strong coordinated dedication to their deployment. We show that carbon pricing covering all world regions is a necessary but insufficient component for the success of mitigation action in order to break the current fossil fuel technology lock-in, unless the price is very high. 

As has been shown earlier \citep{Barker2010}, in the disequilibrium perspective, reductions in power sector emissions may not necessarily imply significant macroeconomic costs (direct and indirect) but instead, could  generate additional industrial activity and employment.

\section{Material and methods \label{sect:Model}}

\subsection{Review of the literature}

The great majority of studies of energy systems at the global scale are made using cost-optimisation computational models (a social planner approach).\footnote{For instance those based on the MARKAL/TIMES/TIAM family of models \citep{MARKAL, ETSAP}, some variants of the MESSAGE model from IIASA \citep{MESSAGE}, the AIM model from NIES in Japan \citep{AIM}, REDGEM70 \citep{Takeshita2011,Takeshita2011b}, DNE21+ \citep{DNE21} and many more.} Meanwhile, the economics of climate change are often represented using general equilibrium economic theory\footnote{Neoclassical, Computable General Equilibrium (CGE), Dynamic Stochastic General Equilibrium (DSGE) and other variations.} \citep[fully rational agent behaviour, perfect foresight and information, carried out by a representative agent, and some variations within these concepts, see ][ for reviews of models]{Loschel2002, IMCP2006, Edenhofer2010}, which tend to yield negative macroeconomic impacts of climate change mitigation, one could argue, by construction.\footnote{For instance in the DICE model \citep{Nordhaus2010}, mitigation costs are subtracted from GDP. In standard CGE models, due to `crowding out', investments equal to those of mitigation are lost to the economy, see section~\ref{sect:Econ}.} In both these approaches, which together represent  the current methodological standard, the assumptions about the nature of agents make the optimisation problem involved tractable. These assumptions about the nature of behaviour, however, may over-simplify aspects of an inherently complex global energy-economic system that are crucial for climate change mitigation, leaving open the question as to how much results stem from these simplifications, and whether relaxing these constraints changes perspectives.

Cost-optimisation technology models, in normative mode, are still the most powerful tools for finding detailed, lowest-cost future technology pathways that reach particular objectives. If used for descriptive purposes, they imply a description of agents  (investors, consumers) as identical, who possess a degree of information and technology access as well as foresight sufficient to generate pathways that are cost-optimal at the system level, which alternatively corresponds to a controlled degree of coordination between all actors involved in the evolution of the system.\footnote{Minimising total system cost as opposed to minimising individual project costs, the level where decisions take place.} As stated in the Global Energy Assessment \citep[ch.~17,][]{GEA2012ch17}, ``\emph{A fundamental assumption underlying the pathways is that the coordination required to reach the multiple objectives simultaneously can be achieved}". While this approach generates a significant simplification to a highly complex system, it may be argued that such a spontaneous emergence of coordination is somewhat unlikely. For instance liberalised energy markets involve actors free to take their investment and consumption decisions based on their particular circumstances, and are only \emph{incentivised} by policy. Thus while optimisation frameworks are valuable for identifying feasible and cost-effective pathways that reach particular objectives at the system level, they do not suggest how exactly to achieve them from a policy standpoint, because they do not specifically model decision making by diverse agents. Strong coordination is difficult to generate from economic policy instruments, leading to sub-optimal outcomes and \emph{technology lock-ins}. 

Meanwhile, equilibrium economic theory implies that climate change mitigation costs are borne at the expense of consumption or investment elsewhere,\footnote{Since investment resources are assumed to be used optimally (full employment), new investment in mitigation `crowds-out' investment elsewhere.} leading to detrimental economic impacts, which is not universally agreed to occur \citep{Barker2010,Grubb2014}. In particular, equilibrium theory relies on decreasing or constant returns to scale, becoming unstable in the presence of processes with increasing returns such as induced technological change \citep{Arthur1989}. Increasing returns also imply the property of \emph{path dependence} and involve complexity, where new \emph{ordering principles} can emerge from the interactions between system parts \citep{Arrow1995, Anderson1972}. Since path dependent systems may not return to equilibrium after disturbances, scenarios diverge from each other for small differences of starting parameters, in a similar way to physical models of the climate. 

Equilibrium economic analyses recommend carbon pricing as the single most efficient policy tool to fix the climate market failure, when equated to the social cost of emitting carbon. However, it is recognised that some new technologies might not successfully bridge the technology innovation `valley of death' to the marketplace at politically practicable carbon prices without further government support \citep{Grubb2014, Murphy2003}. Deriving future scenarios using normative models and equilibrium economics is conceptually inconsistent with the simulation approach of climate science, and hence potentially misleading for many important stakeholders. 

A simulation approach without systems optimisation is possible using known technology dynamics and a  model for decision-making at the firm level by diverse agents \citep{Mercure2012}. Empirically repeatable dynamics are known to exist in scaling up technology systems (e.g. $S$-shaped diffusion), and their costs (learning curves), which have been extensively studied for decades \citep[see the summary and analyses by ][]{Wilson2011,Grubler1999, Grubler1998}. Declining costs of technology with cumulative experience in scaling them up can be modelled either at the bottom-up scale (learning curves, e.g. \citealt{McDonald2001}, criticised by \citealt{Nordhaus2014}), or at the aggregate scale \citep[induced technical change,][]{IMCP2006}, aspects reviewed by \cite{Loschel2002}. Meanwhile, mathematical generalisations of diffusion dynamics have been suggested \citep[e.g.][]{Saviotti1995,Safarzynska2010, Mercure2012, Karmeshu1985, Bhargava1989, Grubler1990, Metcalfe2004} involving dynamic differential equations similar to those in population growth mathematical ecology \citep[i.e. Lotka-Volterra systems, see][]{Lotka1925,Volterra1939,Kot2001} and demography \citep[e.g.][]{Lotka1925, Keyfitz1977}. In combination with evolutionary dynamics \citep[e.g.][]{Nelson1982,Saviotti1991} and evolutionary game theory \citep{Hofbauer1998}, this offers an artillery of powerful concepts at the core of evolutionary economics \cite[for a discussion, see][]{Hodgson2012}. This emphasises innovation, diffusion and speciation\footnote{Speciation in evolutionary theory means increasing diversity as species increasingly subdivide into sub-species through mutation, adapting to changing conditions.} as a source of economic development and growth \citep{Metcalfe1988, Schumpeter1934,Schumpeter1942,Nelson1982}, the clustering of which is possibly responsible for `Kondratiev cycles' \citep{Freeman2001}. These are broadly consistent with the `multi-level perspective' on technology transitions described by \cite{Geels2002,Geels2005}. Furthermore, complexity and path dependence emerge as key concepts to envisage technology dynamics \citep{Silverberg1988b,Dosi1991}. 

Intermediate scale models do exist that introduce investor/consumer diversity in technology choices driving changes in energy supply, end-use and emissions. For example, agent-based models can be used to represent the multi-level perspective on technology transitions \citep{Kohler2009}. Meanwhile models using multinomial logit structures parameterised by survey data provide a natural representation of diversity \citep[e.g. the CIMS model, ][]{Rivers2006,Mau2008,Axsen2009} based on discrete choice theory \citep{Ben-Akiva1985}. Both approaches provide an appropriate replacement for the neoclassical representative agent. However, even models with detailed behavioural treatments do not currently have a complete representation of empirically known technology diffusion patterns as arise for instance in energy systems \citep{Marchetti1978}, which stem from \emph{both} the diversity of choice \citep[e.g. early, middle and late adopters, ][]{Rogers2010} and industrial dynamics \citep[i.e. industrial capacity growth and decline, e.g.][]{Wilson2012b, Grubler1999}. Conversely, evolutionary models of technology innovation-diffusion \citep[e.g. ][]{Safarzynska2012} do not have detailed representations of consumer choice and diversity. But critically, both principles have not yet diffused widely into mainstream global scale integrated climate-energy systems modelling used for informing climate policy \citep[where many models use exogenous diffusion rates, as discussed by ][]{Wilson2013}. Including these would generate an improved methodological paradigm \citep[i.e. the paradigm suggested by][]{Wilson2011,Grubler1999}, which we propose here. In such a framework, optimisation as a source of dynamic force and representative agent behaviour is replaced by empirically known innovation-selection-diffusion dynamics with behaviour diversity, which are not characterised by equilibria, but feature complex dynamics.

\subsection{Technology diffusion in FTT:Power\label{sect:Diffusion}}

\begin{figure*}[t]
\beq
{Technology \choose substitution}_{j\rightarrow i} = {Production \choose capacity \, share}_i\times{Investor \choose choices}_{ij}\times{Technology \choose decommissions}_j \nn
\eeq
		\begin{center}
			\includegraphics[width=1.2\columnwidth]{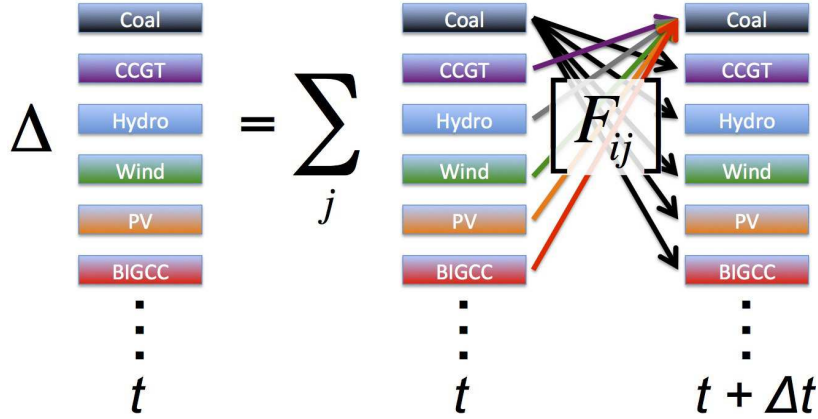}	
		\end{center}
		
		\caption{\emph{Top} Flow of market shares from technology type $j$ towards type $i$. Read from right to left: in a unit of time, out of all decommissions of technology $j$, a proportion is chosen by investors to be replaced by technology $i$, of which only a fraction can be built given the share of production capacity that exists for this technology, restricting the number of units of $j$ that will successfully be replaced by units of $i$. \emph{Bottom} Schematic representation of eq.~\ref{eq:Shares}, where changes in market shares of technologies are equal to the sum of \emph{flows} of shares between categories.}
		\label{fig:shareseq}

\end{figure*}

Emissions reductions in the energy sector can occur through technology substitution, between technologies that produce the same substitutable service (e.g. electricity, heat, etc), through behaviour and practice changes and through reductions in the consumption of that service altogether. Technological change occurs primarily at the average rate of replacement of existing technology as it ages, which is inversely related to its life span. However, notwithstanding lifetime considerations, the number of new units of technology of a particular type that can be constructed at any one time can be larger if that industry is in a well established position in the marketplace, with a large production capacity, than if it is emerging. Even when an emerging technology is (or is made) very affordable, it may not always be accessible to every investor making a choice between available options. Thus in the analysis of the diffusion capacity of technologies, not only cost considerations as seen by diverse agents come into play, but also a limited access to technology and information, and these principles form the core of FTT:Power (Fig.~\ref{fig:shareseq}, see \citealt{Mercure2012} for a general model description, and \citealt{Mercure2013b} for a detailed mathematical derivation).

\ref{app:shares} provides a mathematical derivation of the technology dynamics at the heart of the FTT model, given here. Using the variable $S_i$ for the generation capacity market share of a technology, the rate at which shares of one technology type ($j$) can be replaced by shares of another type ($i$) is proportional to:
\begin{enumerate}
\item The rate at which units of technology $j$ come to the end of their working life, with death rate $\tau_j^{-1}$, 
\item How many old units of $j$ require replacement, a fraction $S_j/\tau_j$ of the total share of replacements. 
\item The rate at which the construction capacity for technology $i$ can be expanded, with growth rate $t_i^{-1}$, 
\item The market position of technology $i$, its share of the market $S_i/t_i$.
\end{enumerate}
This implies the following dynamical equation,
\beq
\Delta S_i = \sum_j S_i S_j \left(A_{ij} F_{ij} G_{ij}-A_{ji} F_{ji} G_{ji} \right){1 \over \overline{\tau}} \Delta t,
\label{eq:Shares}
\eeq
with matrix elements $A_{ij}$ expressing the rate of technology diffusion from industrial dynamics, $F_{ij}$ expressing the probability of investor preferences and $G_{ij}$ providing technical system constraints. $\overline{\tau}^{-1}$ is the average sectoral rate of technology turnover. This equation solves to the classic logistic function of time in the special case of two interacting technologies, with diffusion rate equal to $A_{ij} F_{ij} G_{ij}-A_{ji} F_{ji} G_{ji}$. But generally, it is complex and non-linear, and generates slow uptakes at small penetration, then fast diffusion at intermediate penetration, before a saturation near full penetration, the three matrices together determining the pace of change. This complex system cannot be solved analytically but is straightforward to evaluate numerically using time steps. It corresponds to the replicator dynamics equation in evolutionary theory \citep{Hofbauer1998}, equivalent to a Lotka-Volterra set of equations of population dynamics for competing species, also used in mathematical genetics as well as in evolutionary economics \citep{Saviotti1995}: an \emph{ordering principle} emerging from technology interactions.

\subsection{Learning-by-doing and path dependence \label{sect:PathDep}}

Profit-seeking investor choices $F_{ij}$ are driven by cost differences, and these decrease over time as technologies diffuse and follow learning curves \citep[e.g. ][]{McDonald2001, IEALearning}, generating increasing returns to adoption. Learning-by-doing cost reductions stem from the accumulation of technical knowledge on production and economies of scale in expansion of productive capacity for specific technologies. Technology costs are taken here to apply in globalised firms and markets. Emerging technologies have fast cost reductions (e.g. solar panels) while established systems see very little change (e.g. coal plants). Cost reductions are decreasing functions of cumulative investment, not time, and they do not occur if no investment is made. Learning thus interacts with diffusion where it incentivises further uptake, which generates further learning and so on, a highly self-propagating effect which can lead to sudden technology avalanches. 

Such increasing returns to adoption give the crucial property of \emph{path dependence} to FTT:Power \citep{Arthur1989}. As technologies diffuse following investor choices, the full landscape of technology costs continuously changes, and investor preferences thus change. These changes are permanent and determined by past investments, and therefore by the full history of the market, and different futures emerge, depending on investment and policy choices along the way. Technology costs and learning rates are given in \cite{Mercure2012}, with more detail on the 4CMR website.

\subsection{Natural resource use \label{sect:CSC}}

The diffusion of power systems can only occur in areas where energy resources are available, for instance windy areas for wind power, or natural water basins and rivers for hydroelectric dams. Higher productivity sites offer lower costs of electricity production, and tend to be chosen first by developers. Assuming this, the progression of renewable energy systems development follows increasing marginal costs of production for potential new systems as only resources of ever lower productivity are left to use (decreasing returns to adoption). This is well described by cost-supply curves \citep[e.g. as in][]{Hoogwijk2004,Hoogwijk2009}. For this purpose, a complete set of curves was previously estimated from combined literature and data for 190 countries and 9 types of renewable resources \citep{MercureSalas2012}. These were aggregated for the 21 regions of E3MG (\ref{app:class}). This produced 189 cost-supply curves that are used to constrain the expansion of renewable systems in FTT:Power. The consumption of non-renewable resources is however better represented using a depletion algorithm, described next. 

\subsection{Fossil fuel cost dynamics \label{sect:InverseP}}

Non-renewable energy resources lying in geological formations have an arbitrary value that depends on their cost of extraction, but also on the dynamics of the market. To their cost of extraction is associated a minimum value that the price of the commodity must take in order for the extraction to be profitable. These costs are however distributed over a wide range depending on the nature of the geology (e.g. tar sands, ultra-deep offshore, shale oil and gas, etc). Thus, given a certain demand for the commodity, the price is a function of the extraction cost of the most expensive resource extracted in order to supply the demand, and it separates what is considered \emph{reserves} from \emph{resources}. As reserves are gradually consumed, the marginal cost increases generating a commodity price increase that \emph{unlocks} the exploitation of resources situated in ever more difficult locations with higher extraction costs. For example, tar sands became economical and saw massive expansion above a threshold oil price of around 85-95\$/boe \citep{NEB2011}.\footnote{Including upgrading costs. Extraction costs may have changed since and significant uncertainties exist around these values, which are allowed for in the model.} Thus, to any commodity demand path in time will correspond a path dependent commodity price. The algorithm used here is described with an analysis in \cite{MercureSalas2013b}, relying on data from \cite{MercureSalas2012}. In FTT:Power, this model is used to determine fuel costs for fossil fuel and nuclear based power technologies in global markets.



\subsection{Modelling the global economy: E3MG \label{sect:E3MG}}

E3MG (and variant E3ME\footnote{For details of the econometric equations in both models, see the E3ME website and manual at \burl{www.e3me.com}.}) is an out-of-equilibrium macroeconometric model of the global economy that has been used widely for studies of climate change mitigation macroeconomics \cite[e.g.][]{Barker2012, Barker2010, Barker2006}. It evaluates the parameters of 28 econometric equations using data from 1971 to 2010, and extrapolates these equations between 2010 and 2050. The model features a high resolution: its equations are evaluated for 21 regions of the world, 43 industrial sectors, 28 sectors of consumption, 22 fuel users, and 12 fuel types. Sectors are interrelated with dynamic input-output tables. The model does not optimise economic resources but incorporates endogenous growth and endogenous technical change. This is done following Kaldor's theory of cumulative causation (\citealt{Kaldor1957, Lee1990}, see also \citealt{Loschel2002}), where Technology Progress Indicators (TPIs) are created by cumulating past investments $I$ and R\&D spending using a relative time-weighting,\footnote{$\tau_1$ plays the role of knowledge depreciation while $\tau_2$ is a weighting parameter for R\&D.} 
\beq
TPI(t) \propto \sum_{a = 0}^\infty e^{-a \tau_1} \ln \left(I(t-a) + \tau_2 R\&D(t-a)\right).
\label{eq:TPI}
\eeq
Such TPIs are used in the industrial prices, international trade and employment regressions. Lower prices incentivises higher consumption, and thus industrial investment and R\&D expenditures for production capacity expansions, which lead to lower prices and so on, producing a self-reinforcing cycle of \emph{cumulative causation of knowledge accumulation}. Including accumulated investment and R\&D makes E3MG non-linear, path-dependent and hysteretic, and thus far from equilibrium. E3MG region definitions are given in \ref{app:class}.

\subsection{Endogenous technical change and energy price-demand interactions \label{sect:E3MG}}

The demand for electricity depends on its price, and it is well known that in situations of high electricity prices, people may strive to find more effective ways to use their income, preferring to invest in more efficient technology, perceived as a worthwhile tradeoff, or to simply reduce their consumption. When the electricity supply technology mix changes, the minimum price at which electricity can be profitably sold also changes, and with such price changes, the demand for electricity changes. For example, when carbon pricing or feed-in tariffs are used to ensure access of expensive renewables into the grid, the price of electricity increases, affecting consumer demand and behaviour. Thus, reductions in emissions originate from \emph{both} a change in the carbon intensity of the power sector and changes in the demand for electricity. These aspects of energy economics are prominent in this work, responsible for a significant fraction of our calculated emissions reductions in scenarios of climate policy.

Electricity demand is modelled in E3MG using an econometric equation that incorporates a contribution from spillovers from investment and R\&D spending in other sectors (\citealt{Forssell2000}, see also the E3ME manual, \citealt{E3MEManual}). Since new investments tend to involve technologies with higher energy efficiencies and because the turnover of capital does not allow return to old technology, here the TPI is formed by cumulating positive increases in investment and R\&D, which thus cannot decrease.\footnote{According to the study performed by \cite{Forssell2000}, this describes historical data better than the TPI given in eq.~\ref{eq:TPI}.} The energy demand equation takes the form
\beq
X_{ik} = \beta^0_{ik} + \beta^1_{ik}Y_{ik} + \beta^2_{ik} P_{ik} + \beta^3_{ik} TPI_{ik} + \epsilon_{ik},
\eeq
where for fuel $i$ and region $k$, $X_{ik}$ is (in log space) the fuel demand, $Y_{ik}$ represents sectoral output, $P_{ik}$ relative prices. Economic feedbacks between FTT:Power and E3MG occur with four quantities: electricity prices, fuel use, power technology investments and tax revenue recycling. 

\subsection{Emulating large models of the natural world}

Detailed models of the Earth system are highly dynamical, complex and computationally demanding. An efficient way to integrate dynamic responses to inputs from other models is to create reduced-order statistical representations of model outputs (`emulators') which can be used as surrogate models for coupling applications. Emulators provide a method of analysing the otherwise intractable cascade of uncertainty across multiple complex systems. This approach was used here in order to obtain representations of the planet's carbon cycle and its climate system, by emulating data produced by the large models PLASIM-ENTS and GENIE-1 \citep{Foley2014}. \ref{app:climate} provides a detailed account of the procedure. 

E3MG-FTT emissions in each scenario were fed to the carbon cycle emulator, in order to obtain a trace of CO$_2$ concentrations within an uncertainty range. This trace with uncertainty was fed to the emulator of the climate system in order to obtain a trace of future global warming and climate change within an uncertainty range, which thus cascades the uncertainty of both models.

\section{Results and discussion \label{sect:Proj}}
\subsection{Scenario creation and policy instruments \label{GlobalProj}}

\begin{figure*}[p]
		\begin{center}
			\includegraphics[width=2\columnwidth]{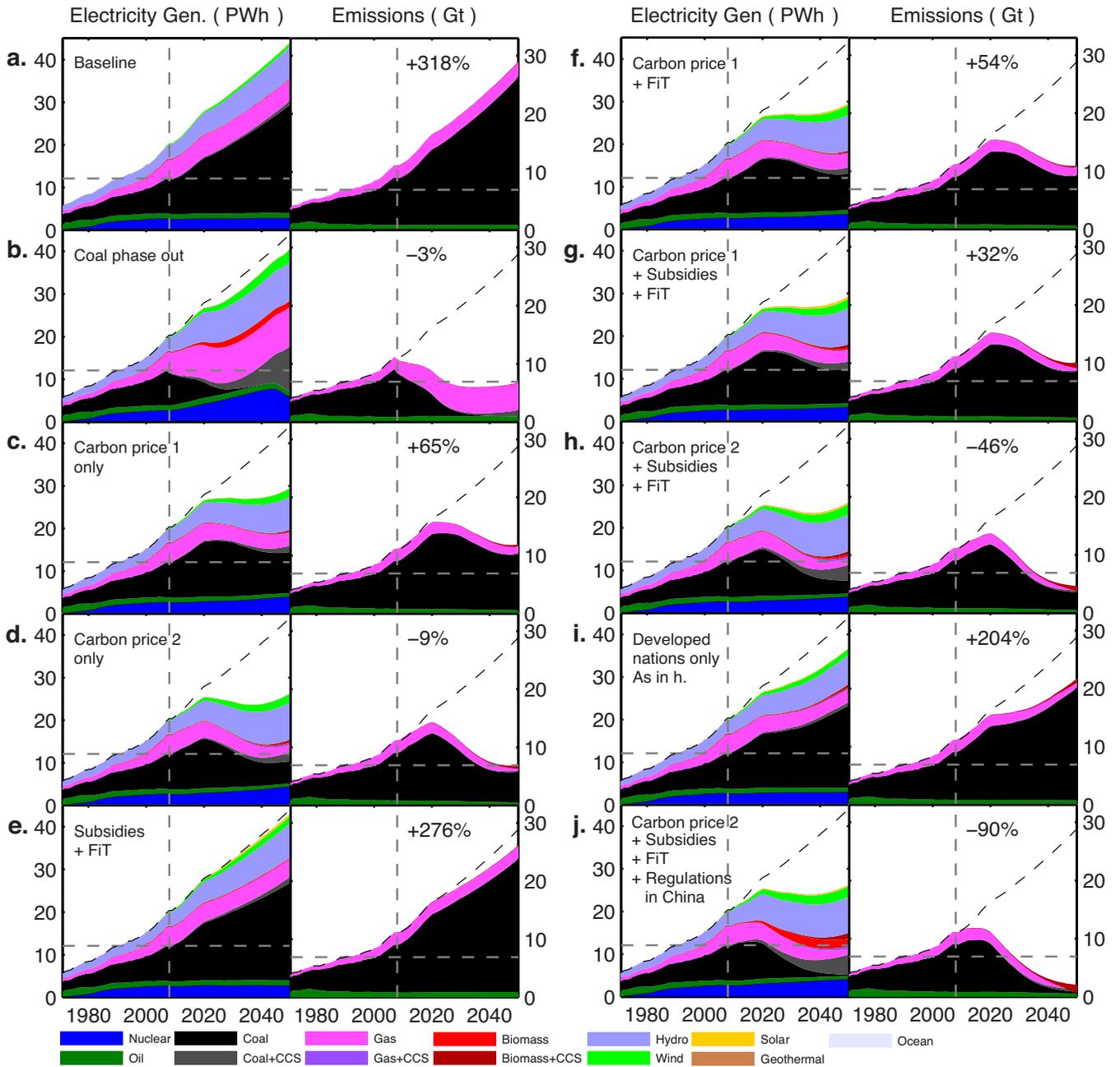}	
		\end{center}
	\caption{Electricity generation and emissions in ten scenarios of policy. The dashed vertical lines delimitate historical IEA data from FTT:Power-E3MG scenarios, while the horizontal dashed lines indicate the 1990 electricity generation and emissions levels. Percent values indicate emissions reductions with respect to the 1990 level. The left panels show electricity generation by technology, while the right panels present the associated CO$_2$ emissions. Red areas of emissions by Biomass+CCS are negative contributions of sequestrated emissions, \emph{reducing} global emissions. FiT indicates a Feed-in Tariff.}
	\label{fig:Figure1}
\end{figure*}

Ten scenarios of electricity policy assumptions  of different types and resulting technology mix and emissions up to year 2050 were created with FTT:Power-E3MG, lettered \emph{a} to \emph{j} (Fig.~\ref{fig:Figure1}). These all lead to different futures for the global power sector and different CO$_2$ emission profiles. It is impractical to reproduce all the information of these simulations in this paper, and therefore a summary of the  results is given here, the details having been made available on the 4CMR website,\footnote{\burl{http://www.4cmr.group.cam.ac.uk/research/FTT/fttviewer}} where they can be displayed in terms of the full resolution of 21 world regions and 24 power technologies, for policy assumptions, electricity generation, emissions and levelised costs. Four energy policy tools were explored: carbon pricing/taxing,\footnote{Carbon pricing or taxing is not conceptually different in FTT, since the price/rate is determined outside of the model. E3MG does not currently solve an endogenous carbon price; it is fully exogenous.} technology subsidies, feed-in tariffs (FiTs) and direct regulations. Individual tools and various combinations were explored, a summary given in figure~\ref{fig:Figure1}. By gradually elaborating various policy frameworks, a scenario was found where power sector emissions are reduced by 90\% below the 1990 level, involving all four policy instruments used simultaneously. Emissions are fed to the carbon cycle and climate model emulators GENIEem and PLASIM-ENTSem in order to determine the resulting atmospheric CO$_2$ concentration and average global warming, for these scenarios where other sectors are not targeted by climate policy.  

The nature of FiTs here is that access to the grid at a competitive price is ensured (a price higher than the consumer price), the difference being paid by the grid and passed on to consumers through the price of electricity.\footnote{FiT prices, although they could be made so, do not depend on capacity here by design.} The consumer price of electricity is raised by just the amount that makes this economically viable. The consumer price in the model is derived from an averaged Levelised Cost of Electricity (LCOE),\footnote{Using historical data, in order to preserve local taxation schemes, scaled to change according to the rate of change of this technology average. For an exact definition of our use of the LCOE, see \cite{Mercure2012} or \cite{IEAProjCosts}.} 
\beq
P \propto \sum G_i LCOE_i/\sum G_i,
\eeq  
with $G_i$ the electricity generation. The LCOE as perceived by investors when a FiT exists includes an `effective subsidy' given by the grid that covers the difference between the levelised cost of a technology and the consumer price of electricity plus a margin (investors here may be corporate or homeowners). In the case of carbon pricing, the LCOE calculation that investors are assumed to perform includes a carbon cost component, and the price of carbon is passed on to consumers through the price of electricity. Thus the price of electricity also increases with the carbon price unless emitting technologies are phased out. 

Technology subsidies are fractions of the capital costs of low carbon technologies that are paid by the government, reducing the LCOE that investors face. These are defined exogenously for every year up to 2050 and are phased out before then, after which it is hoped that the technology cost landscape becomes permanently altered such that technologies do not need to be indefinitely subsidised. Regulations refer to controlling the construction of new units of particular technologies, and can be used to phase out particular types of systems. When a regulation is applied to a technology category, no new units are built but existing ones are left to operate until the end of their lifetime.

Fig.~\ref{fig:Figure1} summarises the result of the policy tools exploration. Electricity generation by technology type is given in the series of panels to the left of each pair, while emissions are given on the right. The vertical dashed lines indicate the start of the simulations in 2008, and trends to the left of this line are historical data from the \cite{IEAWEB2012,IEACO22012}. The horizontal dashed lines indicate the 1990 levels of electricity demand and emissions. Dashed curves correspond to the baseline values for comparison. In all scenarios excluding the baseline, policy schemes generate both a reduction of electricity consumption and emissions. Consumption reduces due to increases in the price of electricity, through the energy demand econometric equation of E3MG, which contributes significantly to emissions reductions. All additional emissions reductions are due to changes in fuels used associated to changes of technologies. CO$_2$ emission levels in 2050 with respect to the 1990 level are given in percent values. 

\subsection{Climate policy for achieving 90\% reductions in power sector emissions \label{GlobalProj}}

The baseline scenario (Fig.~\ref{fig:Figure1} panel \emph{a}), which involves maintaining current policies until 2050,\footnote{Carbon pricing for the EU-ETS only, reaching 80~2008\$/tCO$_2$ in 2050.} leads to global power sector emissions in 2050 of 30~GtCO$_2$/y, 318\% above the 1990 level, and total emissions of 65~GtCO$_2$/y. Cumulative emissions for the time span 2000-2050 amount to 2321~GtCO$_2$. According to the model, this pathway is likely to commit the planet to a warming that exceeds 4$^{\circ}$C above pre-industrial levels in around 2100 (Fig.~\ref{fig:Figure4} and section~\ref{sect:Clim} below, with high probability), consistent with \cite{Zickfeld2009} and \cite{Meinshausen2009}.\footnote{Extrapolating this emissions trend linearly to 2100, where cumulative emissions in 2100 could be $> 9000$~GtCO$_2$, or 2500~PgC, leads to at least a 75\% probability of exceeding 4$^{\circ}$C, especially if emissions do not stabilise by then, according to \cite{Zickfeld2009}. We find a 5\% probability of warming of less than 3.6$^{\circ}$C, see below.} In view of finding ways to reduce the power sector's share of these emissions and to limit global warming, we searched areas of policy space for effective abatement in the short time span. 

\begin{figure}[t]
		\begin{center}
			\includegraphics[width=1\columnwidth]{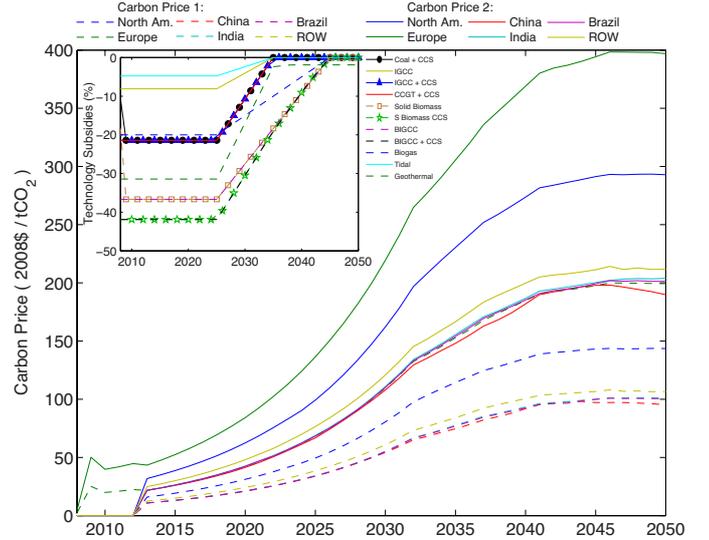}	
		\end{center}
		\caption{(\emph{Main Graph}) Two sets of exogenous carbon prices for all regions of the world, where the set \emph{carbon price 1} was used in scenarios \emph{c,e,f} of figure \ref{fig:Figure1} while the set \emph{carbon price 2} was used in scenarios \emph{c,g,h,i}. (\emph{Inset}) World average of all sets of technology subsidies used in scenarios \emph{e,g,h,j}.}
		\label{fig:CarbonPrices}
\end{figure}

The first option explored (panel \emph{b}) was to use regulations to prevent the construction of new coal power plants worldwide, the systems with highest emissions ($\simeq$1~ktCO$_2$/GWh), unless they are equipped with carbon capture and storage (CCS). This results mostly in a transfer from a coal lock-in to a gas lock-in, reducing global emissions approximately to the 1990 level, largely insufficient for meeting the 2$^{\circ}$C target.

The second option was to use carbon pricing as a unique tool, with different price values for different regions covering all world regions shown in Fig.~\ref{fig:CarbonPrices}, between 100 and 200~2008\$/tCO$_2$ in 2050 (panel~\emph{c}) and between 200 and 400~2008\$/tCO$_2$ (panel~\emph{d}). This measure, mostly generating reductions in electricity consumption due to higher electricity prices, yields emissions of around 65\% above the 1990 level and to 9\% below the 1990 level, respectively. This modest impact suggests that carbon pricing on its own requires very high carbon prices in order to generate significant reductions, or that it is simply insufficient.\footnote{Standard analyses using marginal abatement cost curves assume an instantaneous implementation of mitigation measures, and thus are able to have carbon prices equal to their cost.} However as we show now, combinations of policies achieve this much more effectively. 

As a first combination of policies, FiTs (wind and solar) and technology subsidies (all other low carbon technologies except wind and solar) were introduced without carbon pricing (panel \emph{e}), of order 30-50\% of capital costs for technology subsidies and feed-in prices 5-15\% above the electricity price for FiTs. The fine details depend on regions and technologies, see the inset of Fig.~\ref{fig:CarbonPrices} for a world average or the data on our website for details. This generates very modest uptakes of low carbon technologies and thus small changes in emissions compared to the baseline, 276\% above 1990. This is due to the low cost of producing electricity using fossil fuels in comparison to all other technologies, in particular coal, and therefore without very high subsidies, carbon pricing is necessary in order to bridge this cost difference. 

Scenario $f$ shows the use of carbon pricing up to 200\$/tCO$_2$ with FiTs, the latter generating very little change over scenario \emph{c}. Using carbon pricings of up to 200\$/tCO$_2$ in combination with technology subsidies and FiTs in all world regions (panel~\emph{g}) yields emissions of 32\% above the 1990 level, still insufficient. With carbon pricing of up to 400\$/tCO$_2$ in combination with the same set of technology subsidies and FiTs (panel~\emph{h}), reductions are much larger, 46\% \emph{below} the 1990 level. This indicates how the impact of policy combinations may be larger than the sum of the impacts of its components taken separately, offering \emph{significant potential synergies}. 

A scenario was explored where only the developed world applies the stringent climate policies of scenario \emph{h} (panel~\emph{i}), in which it is hoped that this generates enough investment to bring the costs of low carbon technologies down into the mainstream, thus becoming accessible to developing or under-developed countries. We see no noticeable uptake of new technology in these countries, costs remaining unaccessible especially in comparison with coal based technologies, and as a consequence global emissions remain at 204\% above the 1990 level.

A significant amount of the remaining power sector emissions in scenario \emph{h} reside in China (79\%), where the lock-in of coal technology is very difficult to break given the near absence of alternatives with the exception of hydroelectricity, which is driven to its natural resource limits. The choice of investors thus needs to be constrained at the expense of having to sell electricity at higher prices. Therefore regulations were introduced in scenario \emph{j} in China that prevent the construction of new coal power stations unless they are equipped with CCS. This additional policy forces additional diversity in the Chinese technology mix, bringing down global emissions to 90\% below the 1990 level without early scrapping. Note that it is possible that under different scenarios of technology subsidies, FiTs and regulations, the carbon price necessary for these emissions reductions could be lower, requiring further investigations in this complex parameter space. Total cumulative emissions for the time period 2000-2050 in scenario \emph{j} (in the baseline) are of 1603~Gt (2321~Gt), given that other sectors do not change their technologies significantly, of which 350~Gt (893~Gt) originate from the electricity sector alone. 

\subsection{Climate change projections \label{sect:Clim}}

\begin{figure}[p]
		\begin{center}
			\includegraphics[width=1\columnwidth]{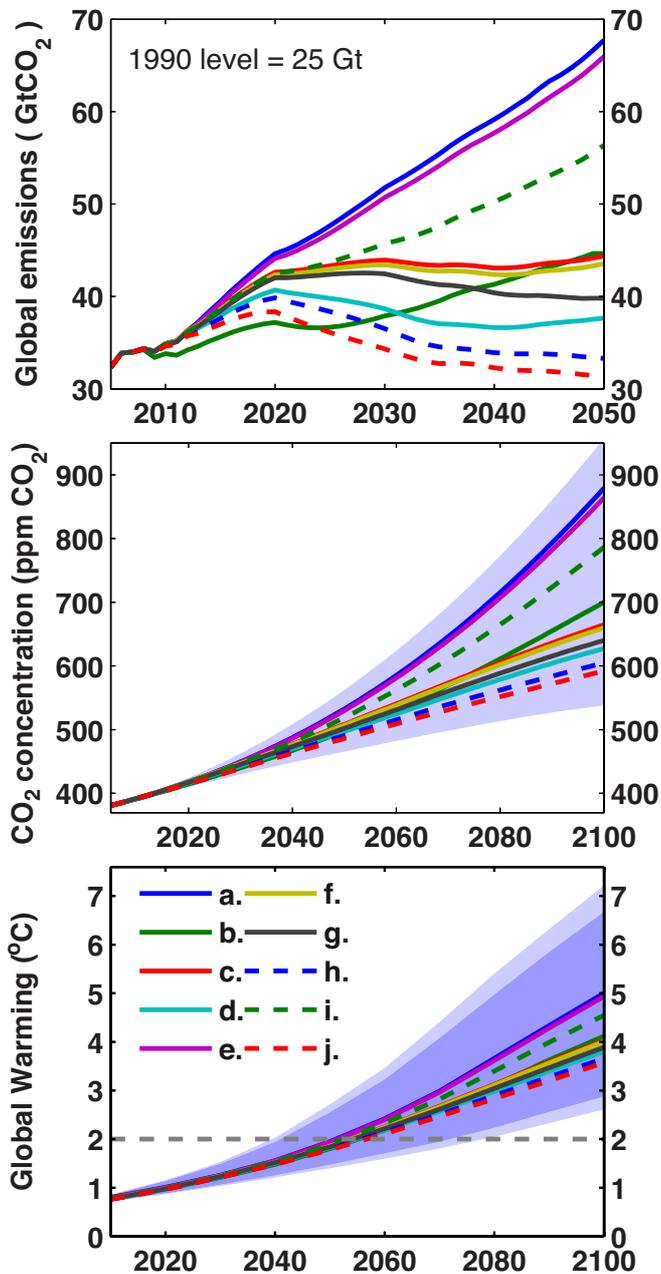}	
		\end{center}
	\caption{(\emph{Top}) Total global anthropogenic emissions in all ten scenarios of Fig.~\ref{fig:Figure1}. (\emph{Middle}) CO$_2$ concentrations, the blue area representing the 95\% confidence range. (\emph{Bottom}) Global warming with respect to the pre-industrial level, the inner blue area delimitating the 95\% confidence range of the climate model, and the outer blue area showing the combined carbon cycle and climate model uncertainties. The top of the blue areas indicate the uncertainty boundary for the highest concentration and warming scenario, while the bottom of the blue area refers to the uncertainty boundary of the lowest concentration and warming scenario.}
	\label{fig:Figure4}
\end{figure}

Global emissions from all sectors in scenarios $a$ to $j$ were fed into the carbon cycle emulator (GENIEem) in order to calculate the resulting CO$_2$ concentrations with their uncertainty range, which themselves were supplied to the climate system emulator (PLASIM-ENTSem) in order to find out their climate change impacts with climate uncertainty, cascading the uncertainty of the carbon cycle into that of the climate system. Figure~\ref{fig:Figure4}, top panel, displays global CO$_2$ emissions for all scenarios of Fig.~\ref{fig:Figure1}, including however all fuel combustion emissions from endogenous sources in the model as well as exogenous trends of emissions for non-fuel-related sectors (e.g. land use), obtained from the EDGAR database. While the changes modelled include those in power sector emissions of Fig.~\ref{fig:Figure1}, they also include modest changes in other sectors (e.g. industry) occurring due to carbon pricing for all fuel users subject to the emissions trading scheme and due to changes in economic activity.

In order to run the climate model emulator, emissions were required up to 2100. In complex models such as E3MG or climate models, uncertainty increases with time span from the present. This currently makes convergence more difficult in E3MG beyond 2050, especially in scenarios of stringent climate policy which may lead the model to venture near the boundary of the behavioural space defined by its econometric relationships prescribed by data prior to 2010. However the primary interest in this work resides in assessing the impacts of near term policy action on the future state of the world. Since the baseline scenario emissions trend is very nearly linear, it was extrapolated with a polynomial to 2100, as well as those of scenarios $e$ and $i$. All other scenarios feature stabilised emissions in 2050 ($b$, $c$, $d$, $f$, $g$, $h$ and $i$), and thus their 2050 emissions values were assumed to be maintained constant up to 2100.


The middle panel of Fig.~\ref{fig:Figure4} shows the resulting atmospheric CO$_2$ concentrations, with uncertainty given as a blue area. It was observed that scenario \emph{a} already reaches a median value of $533\pm30$~ppm in 2050 while scenario \emph{j} reaches $485\pm20$~ppm. This is above the generally agreed threshold of 450~ppm for maintaining warming below 2$^\circ$C. These concentrations with uncertainty were fed to the climate model emulator (Fig.~\ref{fig:Figure4}, bottom). For the baseline scenario, this yielded global warming median temperature changes of between $3.9^\circ$C and $6.7^\circ$C over pre-industrial levels with a median value of $5.0^\circ$C\footnote{A temperature change of $0.6^\circ$C between pre-industrial levels and 2000 was assumed, see \cite{Meinshausen2009} and NASA data at \burl{http://data.giss.nasa.gov/gistemp/tabledata_v3/GLB.Ts+dSST.txt}.} when using the median concentration and only the climate model uncertainty, and between $3.6^\circ$C and $7.2^\circ$C with the same median when cascading the carbon cycle uncertainty into that of the climate model. This therefore could in principle exceed as high as $7.2^\circ$C of warming with a low probability. 

Meanwhile, the electricity decarbonisation scenario $j$ yields warming values of between $2.9^\circ$C and $4.8^\circ$C, median of $3.6^\circ$C, with carbon cycle uncertainty only and between $2.6^\circ$C and $5.2^\circ$C, same median, with both carbon cycle and climate model uncertainties. The electricity sector decarbonisation scenario thus has a negligible probability of not exceeding $2^\circ$C of warming. This indicates that the decarbonisation of the power sector by as much as 90\% is insufficient if other sectors such as transport and industry are not specifically targeted by climate policy, in order to avoid `dangerous' climate change.

\subsection{Learning cost reductions and energy price dynamics \label{sect:Prices}}

\begin{figure*}[t]
		\begin{center}
			\includegraphics[width=2\columnwidth]{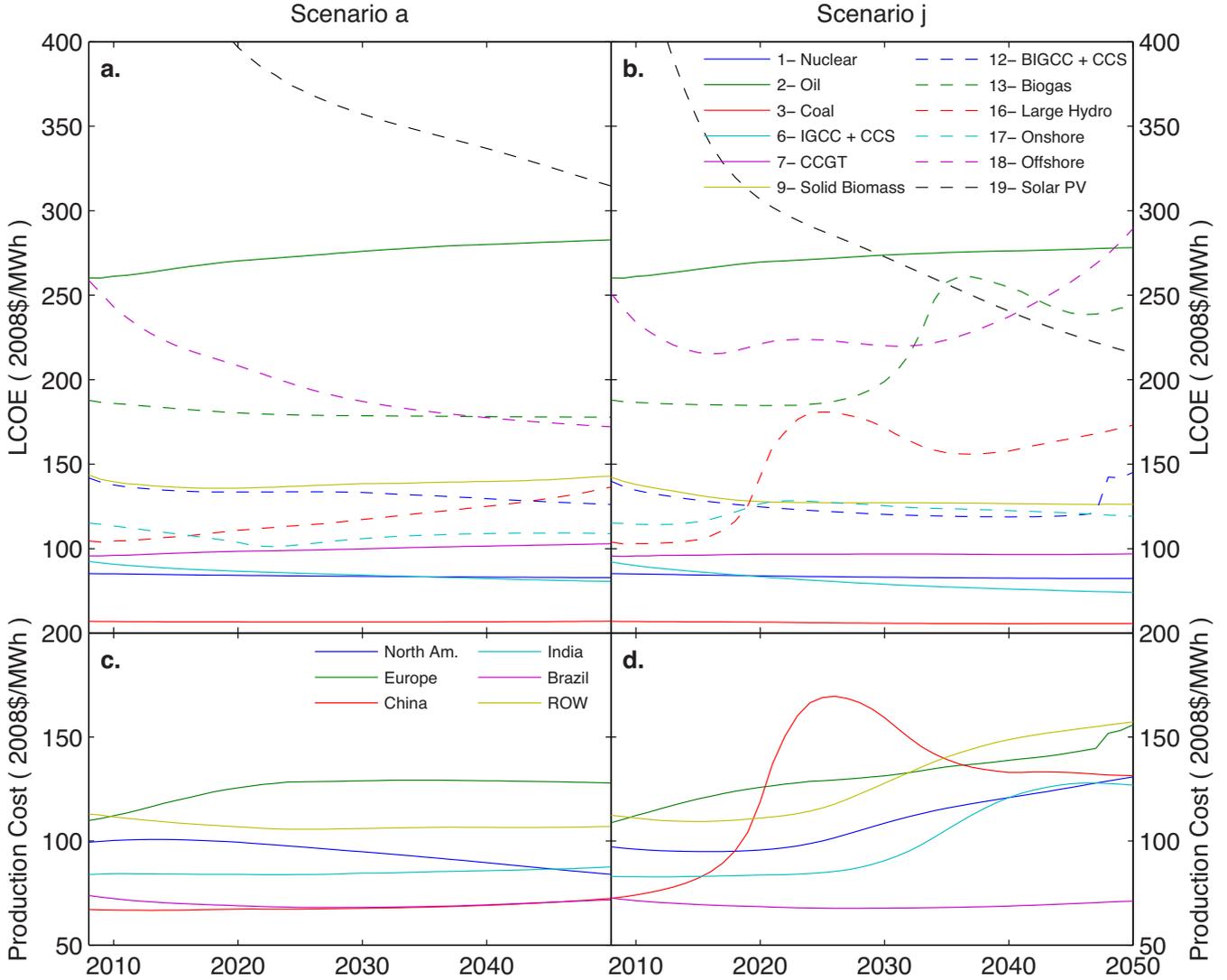}	
		\end{center}
	\caption{(\emph{Top panels}) LCOE per technology excluding carbon pricing, technology subsidies and FiTs, for the baseline (\emph{a}) and the mitigation scenarios (\emph{b}). \emph{Bottom Panels} Marginal cost of electricity production for 6 world regions, for the baseline (\emph{c}) and the mitigation scenarios (\emph{d}).}
	\label{fig:LCOE_Prices}
\end{figure*}

The uptake of low carbon technologies generate learning cost reductions that alter permanently the technology cost landscape. Figure \ref{fig:LCOE_Prices} shows world averages of bare technology costs (upper panels) for the baseline and mitigation scenario \emph{j}, regionally weighted by electricity generation, excluding technology subsidies, the carbon price and FiTs. These values, when including policy, drive investor choices in both the baseline (left) and the mitigation (right) scenarios. Roughly speaking, decreases stem from learning-by-doing cost reductions while increases originate from increasing natural resource scarcity with development. While the cost of PV panels decreases in the baseline scenario mainly due to deployment in Europe, it decreases by more than half its 2008 value in the mitigation scenario where they benefit from FiTs everywhere. Meanwhile, onshore wind power does come into the mainstream in many regions of the world in the mitigation scenario and does not necessitate support all the way to 2050, where the value of the wind FiTs become near zero or even negative, in which case the policy is dropped altogether. In other regions, wind power is limited by resource constrained decreasing capacity factors and corresponding increasing costs. Other technologies, such as geothermal or wave power (not shown), see very little uptake in this particular mitigation scenario and therefore little cost reductions.

The costs of producing electricity, defined as share-weighted average LCOEs, are given for 6 aggregate regions in the lower panels of Fig.~\ref{fig:LCOE_Prices}. Such a marginal cost is used in E3MG to construct electricity prices in 21 regions, of which the changes alter electricity consumption. These are different between regions, stemming from different technology and resource landscapes, where lower marginal costs correspond to higher shares of coal based electricity. Significant increases are observed in the 90\% decarbonisation scenario in all regions, reflecting the cost of the energy transition passed-on to consumers. 

The marginal costs of fossil fuels are calculated using estimates of reserves and resources, described in section \ref{sect:InverseP}, and are not highly affected by changes in policy in these scenarios. In both scenarios oil and gas costs increase significantly up to 2050 in a similar way, but these increases are dampened by the massive accession to unconventional fossil fuel resources (oil sands, heavy oil and shale gas). This analysis will be expanded elsewhere. Coal costs are only moderately affected by changes in demand due to large coal resources. The cost of natural uranium ore is stable until 2035 where an increase is observed, generated by increasing scarcity, and at this level of consumption, U resources are projected to run out before 2100 unless technology changes \citep[e.g. thorium reactors, see ][]{MercureSalas2012, MercureSalas2013b, IAEA2009}.

\subsection{Global economic impacts of a 90\% reduction scenario \label{sect:Econ}}

The macroeconomic impacts of scenarios \emph{a} and \emph{j} in E3MG is a vast subject beyond the scope of the present paper, and will only be summarised here. We find that decarbonising the electricity sector by 90\% has moderate \emph{economic benefits}, generating additional employment, real household income and increases in GDP of between 1 and 3\% (depending on the region) in comparison to scenario \emph{a}, broadly consistent with previous similar analyses performed with the model \citep{Barker2010,Barker2006}. This is due to two opposing forces acting against one another: the introduction of low carbon technologies force increases in electricity prices (as seen in Fig.~\ref{fig:LCOE_Prices}), lowering real household disposable income, while low carbon technology production generates further employment in various industrial sectors, increasing household income. These were observed to roughly cancel each other out, which is possible \emph{as long as labour and capital (investment) resources can be made available} \citep{Barker2010}. 

In our scenarios, carbon pricing generates government income larger than government spending on technology subsidies, the rest being redistributed to households in the form of income tax reductions, increasing further their disposable income. The resulting impacts are therefore of increased household income and consumption in comparison to the baseline and thus higher GDP. It is to be noted however that there are winners and losers in this picture both in terms of sectors and world regions, depending how much they depend on activities of the oil, gas and coal sectors.

In this disequilibrium \emph{demand-led} perspective, our assumptions about capital and labour markets are consistent with our assumptions of energy markets, in that these resources are not assumed currently used optimally and their markets do not automatically produce optimal outcomes. This means that excess finance as well as unemployment exist in the model. This contrasts with the results of many other economic models used to assess the economic impacts of climate change mitigation \citep[e.g. see the model comparison in ][]{Edenhofer2010}, which tend to give moderate negative impacts. The main difference stems precisely from assumptions over economic resources: in general equilibrium theory, displacing economic resources that are optimally used can only lead to effects detrimental to the economy (crowding out effects). However unemployment does exist in the world economy, and it is not generally agreed that investment resources are currently used to their utmost potential \citep{Grubb2014}. Further research into this issue is crucial and requires modelling the global financial sector, absent in equilibrium theory,\footnote{Rational expectations in equilibrium theory mean that agents with perfect foresight \emph{never default} on their debts, which implies a non-existence of risk in finance.} but also not specifically treated in our model.

\subsection{Local scenarios of power generation and emissions \label{sect:Local}}

\begin{figure*}[p]
		\begin{center}
			\includegraphics[width=2\columnwidth]{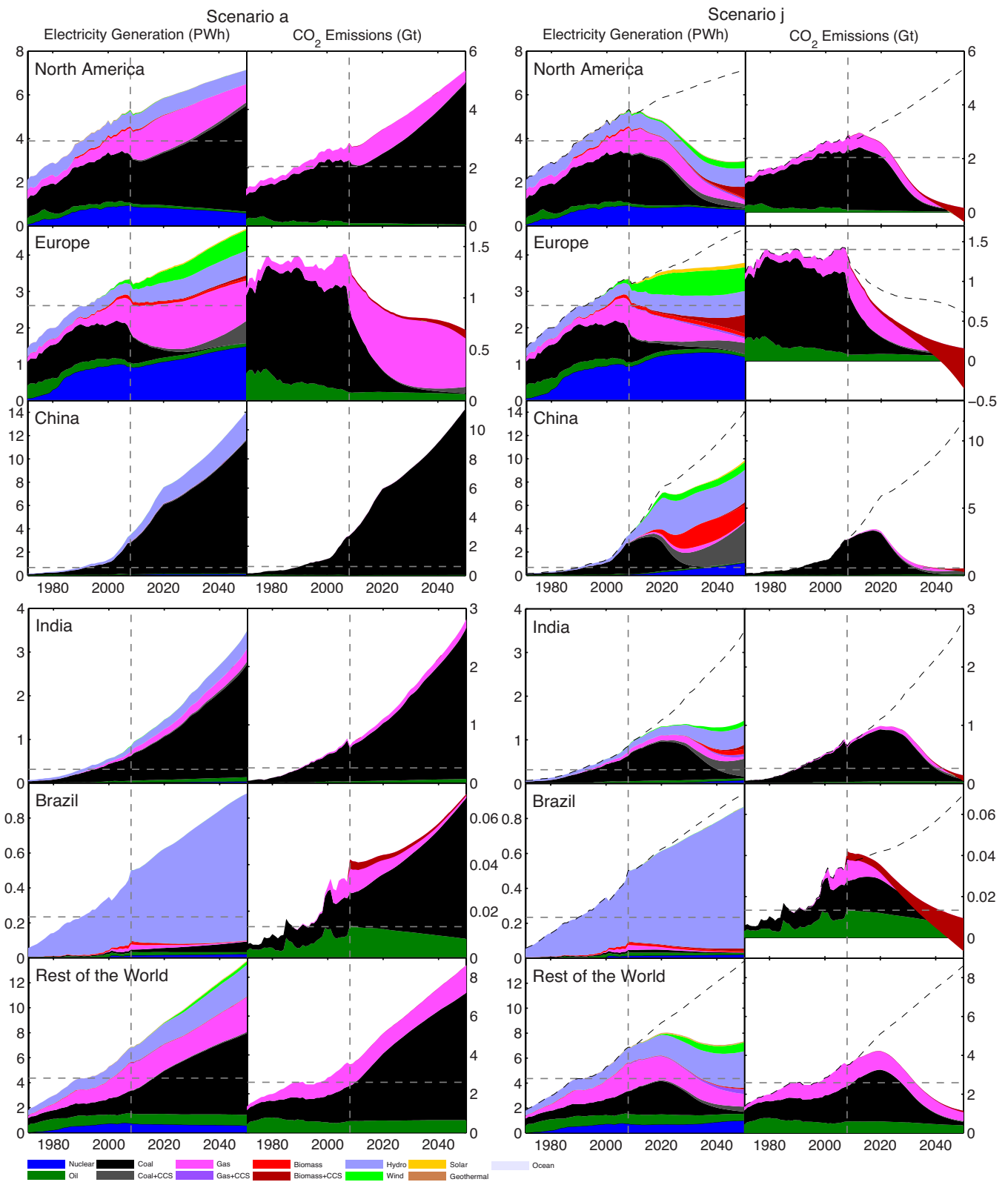}	
		\end{center}
	\caption{Electricity production and CO$_2$ emissions for six regions covering the world, with associated emissions in two scenarios. In all plots, the vertical dashed lines separates historical IEA data from FTT:Power-E3MG scenarios, while horizontal dashed lines indicate the 1990 level of CO$_2$ emissions. Figures on the left are for the baseline scenario while those on the right are for the 90\% mitigation scenario, the sum of the data for all regions being equal to the data of figure \ref{fig:Figure1}. Note that red patches in emissions data, for biomass with CCS, indicate negative contributions.}
	\label{fig:Figure3}
\end{figure*}

It proves instructive to analyse electricity technology landscapes in individual regions of the world in FTT:Power-E3MG, for policy analysis and for better understanding the nature of technology lock-ins and the \emph{restricted} local ability to change in a diffusion perspective. This is presented in Fig.~\ref{fig:Figure3} for six key regions or countries: North America, Europe, China, India, Brazil and the Rest of the World, which have different electricity landscapes stemming from differing energy policy strategies and engineering traditions historically, as well as natural resource endowments. National strategies, reflecting local engineering specialisation related to technology lock-ins, is a natural outcome of this model's structure (eq.~\ref{eq:Shares}), which reproduces the better ability of dominating industries to capture the market despite costs.  

Renewable energy systems are more exploited in Europe than anywhere else in the world, except in Brazil, where hydroelectricity dominates. Europe also sees the most diverse electricity sector, with large amounts of wind power already in the baseline scenario, predominantly in northern Europe and the British Isles, large amounts of nuclear power in France, and some solar power in Germany. Coal fired electricity is mostly phased out before 2050 in the 90\% scenario, resulting in significant emissions reductions. 

North America features higher use of fossil fuels for power production than Europe. However, while E3MG projects a larger potential for consumption reductions, large opportunities for diversification also emerge with significant potentials of renewable energy. Bioenergy with CCS generates a large contribution to American emissions reductions.\footnote{Note that negative emissions from biomass combustion and sequestration could involve a transfer of emissions from the power to the land use sectors, if important land use changes take place, or if fossil fuels are used in the production of biomass. This should be investigated using consistent land-use modelling.} 

China and India have very low technology diversity and important fossil fuel lock-ins. The amount of coal used in China in the baseline is responsible for 10 out of 30~Gt of global emissions in 2050. Diversification proves difficult given the scale of the rate of increase in consumption;  breaking the coal lock-in requires regulations in China to phase out building new coal generators. Large scale diffusion of renewables is slow and retrofitting CCS to coal generators offers a useful alternative. Electricity demand reductions are very large, which requires further investigations for fuel poverty and other social implications.  

In Brazil, even though hydroelectricity is not the least expensive resource, it nevertheless dominates, another form of technology lock-in. This is typical of a national engineering tradition dominated by a technology for decades.\footnote{A similar situation exists in France with nuclear power, in Canada with hydroelectricity, in China with coal-fired power stations, and originates from either or both an abundance of resources and historical energy policy strategies that have shaped the local expertise, becoming a `tradition' despite cost considerations.} Brazil is projected to persist in  developing its hydropower capacity despite higher costs and a decreasing potential, until the cost becomes prohibitively high and only less productive hydro resources remain. 

The rest of the world includes predominantly countries where the diversity of existing technologies is low, and persists in this direction. It features large amounts of oil use for electricity despite high oil prices, due to restricted access to technology or fossil fuel subsidies, which are not successfully phased out despite being the least cost-effective way of producing electricity. Coal based electricity makes the dominant contribution to emissions in the baseline, the rest divided between oil and gas fired power stations, for a total of 12 out of 30~Gt of global emissions in 2050 in the baseline. In the mitigation scenario, a significant additional hydroelectricity potential is developed, and coal is replaced by gas turbines, which are eventually retrofitted with CCS. 

\section{Conclusions and policy implications\label{sect:Policy}}
\subsection{Synergy between policy instruments}

This paper shows that in a coupled energy-economy-environment model that does not assume economic equilibrium or use technology cost-optimisation, the impact of policy instruments can be different if used individually or in combinations: the impact of combined policy packages does not correspond to the sum of the impacts of individual instruments. Thus significant \emph{synergies} exist between policy instruments. In this regard we showed that in a technology diffusion perspective, carbon pricing alone is not likely capable of delivering sufficient emissions reductions unless it is unrealistically high; it requires to be combined with technology subsidies, FiTs and regulations. This can be ascribed largely to the inertia of diffusion, and contrasts with the neoclassical environmental economics view that pricing the externality generates the desired outcome most efficiently \citep[e.g.][]{Nordhaus2010, Tol2013}\footnote{Note that the concept of equating a \emph{marginal abatement cost} to a \emph{social cost of carbon} in a path-dependent perspective is ambiguous: several prices can be assigned to both quantities depending on the context and history.}. Our model results indicate that relying on carbon pricing alone even up to 400~2008\$/tCO$_2$ is likely to lead to a \emph{status quo} in the technology mix while delivering very expensive electricity to consumers. Similarly, technology subsidies and FiTs on their own have little impact unless they are combined with sufficiently high carbon pricing. 

We furthermore suggested that particular combinations of policy instruments can produce such strong synergy that reductions of electricity sector emissions of 90\% by 2050 (61\% of 2000-2050 cumulative baseline power sector emissions) become possible without early scrapping of electricity generation capital. Such strong reductions could be complemented by additional reductions in other emissions intensive sectors with additional cross-sectoral synergies: transport, industry and buildings, warranting further work in this area. If early scrapping is allowed, these reductions could be achieved even faster, but would most likely involve higher costs. Finally, different combinations of the policies analysed here could also lead to 90\% emissions reductions, for example with lower carbon prices and higher technology subsidies, FiTs and/or more regulations.

\subsection{The effect of global knowledge spillovers on technology costs: individual vs global coordinated action \label{sect:GLearning}}

Technology systems typically face a vicious cycle: established technologies thrive because they are established, and emerging technologies see barriers to their diffusion due to the lock-in of established technologies. This is the case unless an emerging technology is a radical improvement over the incumbent, or it benefits from sufficient external support. Emerging technologies require investment and sales in order to benefit from improvements and economies of scale: repetition, trial and error enables entrepreneurs to improve their products. They thus require a continuous flow of funds from sales or external investment in order to survive until their products take off on their own in the market.  In the long run, these investments may or may not generate a return, and are thus risky. Without any investment to bridge the `technology valley of death', however, they may become failed innovations. 

Given estimated learning curves of power systems, a certain additional capacity of emerging technologies such as wind turbines and solar PV panels must be deployed in order to bring down their costs to a competitive level set by incumbent technologies. As we find here, this additional capacity is very large, and cannot be deployed by a single nation such as Germany or even the whole of Europe, for the rest of the world to benefit. In contrast, we find that only a concerted global climate policy effort can bring down costs to manageable levels and bring new power technologies into the mainstream, opening very large renewable energy potentials such as that of solar energy. Such a concerted effort can significantly and permanently alter the global landscape of power technology costs and availability. We stress that all countries of the world can benefit from learning cost reductions that originate from investments and sales occurring in various locations. This problem therefore possesses the features of a classic free-rider and collective action problem, where international coordination is the only way by which these cost reductions can take place. Emerging or developed nations cannot simply `wait' for climate policy in other nations to generate diffusion and enough learning cost reductions for new technologies to become competitive: without their involvement they might potentially never become competitive. If the power sector is to decarbonise by 2050, all countries are most likely required to make a contribution to the development of the renewables industry.

\section*{Acknowledgements}

We acknowledge T. Barker for guidance, D. Crawford-Brown for support, M. Grubb for insightful suggestions and M. Syddall for creating the online data explorer. We thank the students who have attended the Energy Systems Modelling seminars held at 4CMR in 2013, with whom the discussions on technology and evolutionary economics led to the development and enhancement of this work, and staff at Cambridge Econometrics for a welcoming stay. We warmly thank four anonymous referees, whose suggestions helped us significantly improve the manuscript. This work was supported by the Three Guineas Trust (A. M. Foley), Cambridge Econometrics (H. Pollitt and U. Chewpreecha), Conicyt (Comisi\'on Nacional de Investigaci\'on Cient\'ifica y Tecnol\'ogica, Gobierno de Chile) and the Ministerio de Energ\'ia, Gobierno de Chile (P. Salas), the EU Seventh Framework Programme grant agreement No 265170 `ERMITAGE' (N. Edwards and P. Holden) and the UK Engineering and Physical Sciences Research Council, fellowship number EP/K007254/1 (J.-F. Mercure).

\appendix
\section{Substitution dynamics \label{app:shares}}

The decision-making process by diverse agents can be expressed with probabilistic pairwise comparisons of options. Investors do not all face similar situations and do not weigh different aspects in the same way, which cannot possibly be enumerated specifically in a model. However, when dealing with large numbers of instances of decision making, there will be majority trends in investor choices, who may be assumed, if all relevant considerations are quantified into costs, to be \emph{seeking cost minimisation with the goal of profit maximisation for their own respective firms}.\footnote{Profit maximisation at the firm level, rather than cost minimisation at the whole system level, is the crucial difference between this bottom-up investor behaviour approach the common top-down cost-optimisation. See for instance \citealt{Nelson1982}.} Data derived probability distributions may thus be used in order to avoid enumerating the details of all situations faced by investors, and where a particular technology is on average more profitable to use than a second one, there usually exist specific situations where the reverse turns out to be true. This provides a crucial simple representation of \emph{diversity} in decision-making. In this form, this is a binary logit model \citep[see][]{Ben-Akiva1985}.

The statistical trend of investor preferences may be expressed as a matrix $F_{ij}$ expressing the relative fractions of investor choices between two technologies $i$ and $j$ out of a set. For example, if $F_{ij} = 70\%$ and $F_{ji} = 30\%$, then investors faced with these two options would choose 70\% of the time technology $i$ and 30\% of the time technology $j$. This can be derived using probability distributions for technology costs (derived from recent investment data, a form of \emph{revealed preferences}),\footnote{Here, we have used a published survey, \citealt{IEAProjCosts}. More details are given in \citealt{Mercure2012}.} calculating the number of instances where technology type $i$ is seen as less expensive than $j$. In the spirit of discrete choice theory, the probability that technology $j$ is perceived less expensive than technology $i$, and the converse, are
\beq
F_{ij}(\Delta C) = \int_{-\infty}^{\infty} F_j(C - \Delta C) f_i(C) dC,
\eeq
\beq
F_{ji}(\Delta C) = \int_{-\infty}^{\infty} F_i(C + \Delta C) f_j(C) dC,
\eeq
where $F(C)$ is a cumulative cost distribution function while $f(C)$ is a cost distribution density, while $\Delta C$ is an average cost difference. 

Innovation generates new technologies that live in niches that protect them from the wider market. From those niches, in appropriate changes of market conditions, can emerge and diffuse new socio-technical regimes \citep{Geels2002,Geels2005}. After the innovation phase, at the level of diffusion, technologies enter what we termed the `demographic phase' \citep{Mercure2013b}, because the derivation follows standard population dynamics for competing species, as in \cite{Kot2001}. The four building block arguments enumerated in section~\ref{sect:Diffusion} are summarised as follows.

{\bf(1-2)} The rate at which units of technology come to the end of their lifetime stems from survival analysis, where a cumulative probability of failure yields a survival function $\ell_j(a)$ of age $a$. The number of units retired from operation $\delta_j$, and the number remaining $N_j$, relate to how many were built $a$ years in the past $\xi_j(t-a)$, with life expectancy $\tau_j$:
\beq
N_j(t) = \int_0^\infty \xi_j(t-a) \ell_j(a) da,
\eeq
\beq
\delta_j(t) = \int_0^\infty \xi_j(t-a) {d \ell_j(a) \over da} da \simeq {N_j \over \tau_j}, \quad \tau_j = \int_0^\infty \ell_j(a) da.
\eeq

{\bf(3-4)} The rate at which units of technology get produced depends on available production capacity, and production capacity is built out of profits on sales, leading to a virtuous cycle that gradually builds up. A `birth' function $m_i(a)$ can be defined that enables to determine the growth of production capacity $\delta N_i$ from historical sales $\xi_i(t-a)$ with re-investment rate $R_i$:
\beq
\delta N_i(t) = R_i \int_0^\infty \xi_i(t-a) m_i(a) da \simeq {N_j \over t_i},
\eeq
with a growth rate determined by
\beq
t_i = {\tau_i \over R_i \Phi_i}, \quad \Phi_i = \int_0^\infty m_i(a) da,
\eeq
a growth time constant much shorter than the life expectancy ($t_i \ll \tau_i$, \emph{births} occur $R_i \Phi_i$ faster than \emph{deaths}).

Finally, constraints of the power system prevent some types of technology from dominating.\footnote{For instance related to their supply properties or variability (e.g. as with wind power).} These can be expressed with a second matrix $G_{ij}$, stopping investments that lead to stranded or unused assets due to technical problems (grid stability).

Using the equation in Fig.~\ref{fig:shareseq}, and normalising unit numbers into market shares, we calculate the \emph{flow of market shares} from technology of type $j$ towards category $i$:
\beq
\Delta S_{j\rightarrow i} \propto \left({S_i \over \tau_i}\right) \left(F_{ij} G_{ij}\right)\left({S_j \over t_j}\right) \Delta t \propto S_i S_j A_{ij} F_{ij} G_{ij} \Delta t.\nn
\eeq
Since, according to the binary logit, the reverse choice is allowed, there is a simultaneous reverse flow of substitutions,
\beq
\Delta S_{i\rightarrow j} \propto \left({S_j \over \tau_j}\right) \left(F_{ji} G_{ji}\right)\left({S_i \over t_i}\right) \Delta t \propto S_j S_i A_{ji} F_{ji} G_{ji} \Delta t,\nn
\eeq
the net exchange between categories $i$ and $j$ is
\beq
\Delta S_{ij} = S_i S_j \left(A_{ij} F_{ij} G_{ij}  - A_{ji} F_{ji} G_{ji}\right) {\Delta t \over \overline{\tau}}.
\eeq
Adding up flows between all possible technologies $j$ and category $i$ yields the main equation \ref{eq:Shares}.

\section{Model regions\label{app:class}}

\begin{table}[h!]\footnotesize
\begin{center}
		\begin{tabular*}{1\columnwidth}{@{\extracolsep{\fill}} p{.31\columnwidth}|p{.65\columnwidth} }
			\hline
			E3MG-FTT Region & Member countries\\
			\hline
			\hline
			1- USA 			& USA\\
			2- Japan	 	& Japan\\
			3- Germany	& Germany\\
			4- UK		& UK\\
			5- France		& France\\
			6- Italy		& Italy\\
			7- Rest of EU-15 & Austria, Belgium, Danemark, Finland, Greece, Ireland, Luxemburg, Netherlands, Portugal, Spain, Sweden\\
			8- EU-12 		& Czech, Estonia, Cyprus, Latvia, Lithuania, Hungary, Malta, Poland, Slovenia, Slovakia, Bulgaria, Romania\\
			9- Canada  	& Canada\\
			10- Australia	& Australia\\
			11- OECD NES	& Iceland, New Zealand, Norway, Switzerland, Turkey\\
			12- Russia	& Russia\\
			13- Rest of Annex I & Belarus, Croatia, Ukraine\\
			14- China		& China\\
			15- India		& India\\
			16- Mexico	& Mexico\\
			17- Brazil		& Brazil\\
			18- NICs		& South Korea, Malaysia, Philippines, Singapore, Taiwan, Thailand\\
			19- OPEC		& Algeria, Iran, Iraq, Kuwait, Libya, Nigeria, Qatar, Saudi Arabia, UAE, Venezuela\\
			20- Rest of the World & All other countries not specified elsewhere\\
			21- Indonesia & Indonesia\\
			\hline
		\end{tabular*}
	\caption{World regions in E3MG-FTT.}
	\label{tab:Regions}
\end{center}
\end{table}

\section{Climate system emulation \label{app:climate}}

To assess climate impacts requires models of both the carbon cycle and the climate. Here we use simplified, statistically derived representations (emulators) of these two systems as described in more detail by \cite{Foley2014}. No direct feedback from climate to economy is implemented, but our approach does allow for spatially and temporally resolved analysis of climate impacts, including the effects of uncertainty in the climate system.
 
The carbon cycle is represented by GENIEem, an emulator of the Grid Enabled Integrated Earth system model \citep[GENIE-1][]{Holden2013}. It takes as inputs a time series of anthropogenic CO$_2$ emissions produced by FTT:Power-E3MG and a scenario of non-CO$_2$ radiative forcing (including the effects of CH$_4$, N$_2$O, halocarbons, aerosols and O$_3$) based on the `representative concentration pathway' \citep{vanVuuren2011} that most closely matches the baseline used in this study. GENIEem calculates the extent to which CO$_2$ emissions remain in the atmosphere and produces a time series of atmospheric CO$_2$ concentration as output. Uncertainty in the carbon cycle is captured by varying the internal parameters of GENIE-1, resulting in an ensemble of 86 possible future atmospheric CO2 concentration profiles.

GENIEem, in turn, provides inputs to the climate-system model PLASIM-ENTSem \citep{Holden2013b}, an emulator of the PLAnet SIMulator \citep{Fraedrich2012} - coupled to the Efficient Numerical Terrestrial Scheme \citep{Williamson2006}. Non-CO$_2$ forcing is, again, prescribed. Uncertainty is captured by varying PLASIM-ENTS internal parameters, resulting in a 188-member ensemble of decadally averaged seasonal climate variables.

The combination of the two emulators in the context of this work, with combined uncertainty analysis, is described in detail by \cite{Foley2014}. First, PLASIM-ENTSem is forced with the median of the GENIEem ensemble, and the median, 5th and 95th percentiles of warming from the resulting PLASIM-ENTSem ensemble are calculated; these bounds, therefore, reflect warming uncertainty due to parametric uncertainty in the climate model alone.

Next, PLASIM-ENTSem is forced with the 5th percentile CO$_2$ concentration from the GENIEem ensemble. The 5th percentile of warming from the resulting PLASIM-ENTSem ensemble is calculated. Finally, PLASIM-ENTSem is forced with the 95th percentile CO$_2$ concentration from the GENIEem ensemble, and the 95th percentile of warming from the resulting PLASIM-ENTSem ensemble is calculated.  These two simulations form a second set of bounds, reflecting warming uncertainty due to parametric uncertainty in the climate model and the carbon cycle model.
\bibliographystyle{elsarticle-harv}
\bibliography{../../../CamRefs}

\end{document}